\documentclass[10pt,twocolumn,letterpaper]{article}

\usepackage{cvpr}              

\usepackage{graphicx}
\usepackage{amsmath}
\usepackage{amssymb}
\usepackage{booktabs}
\usepackage{cite}
\usepackage{algorithmic}
\usepackage{textcomp}
\usepackage{times}
\usepackage{epsfig}
\usepackage{multirow}
\usepackage{multicol}
\usepackage{tabularx}
\usepackage{fdsymbol}
\usepackage{url}
\usepackage{bm}
\usepackage{graphicx}
\usepackage{diagbox}

\usepackage[accsupp]{axessibility}  

\usepackage[pagebackref,breaklinks,colorlinks]{hyperref}

\usepackage[capitalize]{cleveref}
\crefname{section}{Sec.}{Secs.}
\Crefname{section}{Section}{Sections}
\Crefname{table}{Table}{Tables}
\crefname{table}{Tab.}{Tabs.}

\def\cvprPaperID{7761} 
\def\confName{CVPR}
\def\confYear{2023}

\begin{document}

\title{LVQAC: Lattice Vector Quantization Coupled with Spatially Adaptive Companding for Efficient
Learned Image Compression}

\author{Xi Zhang\\
Shanghai Jiao Tong University\\
{\tt\small zhangxi\_19930818@sjtu.edu.cn}
\and
Xiaolin Wu\\
McMaster University\\
{\tt\small xwu@ece.mcmaster.ca}
}
\maketitle

\begin{abstract}
    Recently, numerous end-to-end optimized image compression neural networks have been developed and proved themselves as leaders in rate-distortion performance.
    The main strength of these learnt compression methods is in powerful nonlinear analysis and synthesis transforms that can be facilitated by deep neural networks.
    However, out of operational expediency, most of these end-to-end methods adopt uniform scalar quantizers rather than vector quantizers, which are information-theoretically optimal.
    In this paper, we present a novel Lattice Vector Quantization scheme coupled with a spatially Adaptive Companding (LVQAC) mapping.
    LVQ can better exploit the inter-feature dependencies than scalar uniform quantization while being computationally almost as simple as the latter.
    Moreover, to improve the adaptability of LVQ to source statistics, we couple a spatially adaptive companding (AC) mapping with LVQ.
    The resulting LVQAC design can be easily embedded into any end-to-end optimized image compression system.
    Extensive experiments demonstrate that for any end-to-end CNN image compression models, replacing uniform quantizer by LVQAC achieves better rate-distortion performance without significantly increasing the model complexity.
    Code is available at: \href{https://github.com/xzhang9308/LVQAC}{\textcolor{magenta}{https://github.com/xzhang9308/LVQAC}}.
  \end{abstract}
  

\section{Introduction}
\label{sec:intro}
In the past five years, the research on end-to-end CNN image compression has made steady progress and led to the birth of a new class of image compression methods
~\cite{Toderici2016_variable,Toderici2017_Full,Balle2017_End,Theis2017_Lossy,Agustsson2017_Soft,Balle2018_Vari,Minnen2018_Joint,Mentzer2018_Cond,ultra,davd,agdl,mdvd,addl,Lee2019_Context,Chen2020_Learned}.
The CNN compression can now match and even exceed the rate-distortion performance of the previous best image compression methods \cite{wallace1992jpeg, skodras2001jpeg, BPG, VVC},
which operate in the traditional paradigm of linear transform, quantization and entropy coding.

The advantages of the CNN approach of data compression come from the nonlinearity of its analysis and synthesis transforms of the autoencoder architecture, the end-to-end joint optimization of the nonlinear transforms, uniform quantization of the latent space and conditional entropy coding (context-based arithmetic coding) of quantized features.

Apparently, using uniform scalar quantizer in the above CNN image compression framework is motivated by operational expediency more than other considerations. Only at very high bit rate uniform quantization can approach the rate-distortion optimality \cite{gersho2012vector}.  It is very difficult to directly adopt and optimize a vector quantizer (VQ) in the end-to-end CNN architecture design for data compression, because VQ is a discrete decision process and it is not compatible with variational backpropagation that is necessary to the end-to-end CNN training. In \cite{Agustsson2017_Soft}, Agustsson et al. tried to circumvent the difficulty by a so-called soft-to-hard vector quantization scheme. Their technique is a soft (continuous) relaxation of discrete computations of VQ and entropy so that their effects can be approximated in the end-to-end training. However, in \cite{Agustsson2017_Soft} the quantization centers are optimized along with the other modules, which make the whole system quite cumbersome and more difficult to train.
In this paper, we p
ropose a novel Lattice Vector Quantization scheme coupled with a spatially Adaptive Companding (LVQAC) mapping. LVQ can better exploit the inter-feature dependencies than scalar uniform quantization while being computationally almost as simple as the latter.  Even if the features to be compressed are statistically independent, LVQ is still a more efficient coding strategy than scalar uniform quantization. This is because the former offers a more efficient covering of high-dimensional space than the latter, as proven by the theory of sphere packings, lattices and groups \cite{Conway1988_Sphere}.
Moreover, to improve the adaptability of LVQ to source statistics, we couple a spatially adaptive companding mapping with LVQ.  The resulting LVQAC design is computationally as simple and as amenable to the end-to-end training of the CNN compression model as in the originally proposed framework of \cite{Balle2017_End}.

Consequently, for any end-to-end CNN image compression models, replacing uniform quantizer by LVQAC achieves better rate-distortion performance without significantly increasing the model complexity; the simpler the context-sensitive entropy model, the greater the performance gain.
For instance, the checkerboard context model~\cite{he2021checkerboard} has recently been proposed to speed up the decoding process, but it hurts the R-D performance slightly. By incorporating LVQAC into the end-to-end compression CNN with a checkerboard context model, we can achieve the same performance as the counterpart CNN using a far more expensive auto-regressive context model.
This is because uniform scalar quantization needs a more complex model to compensate for its coding inefficiency, whereas LVQAC does not.

\section{Overview of Learned Image Compression and Related Work}
Ballé~\textit{et~al.}~\cite{Balle2017_End} pioneered the end-to-end CNN model for image compression.  Their network design is still shaped by the legacy of classical three-step signal compression algorithms; namely, transform, quantization and entropy coding.
The greatest advantage of the CNN approach over the classical one comes from replacing linear transforms of the latter by more powerful and complex nonlinear transforms, which is the distinctive prowess of deep neural networks.
Specifically, in the end-to-end image compression CNNs, an image vector
$\bm{x} \in \mathbb{R}^N$ is mapped to a latent code space via a parametric nonlinear analysis transform, $\bm{y}=g_a(\bm{x}; \bm{\theta_g})$. Then this representation is quantized, yielding a discrete-valued vector $\bm{\hat{y}} = Q(\bm{y}) \in \mathbb{Z}^M$ which can be losslessly compressed using entropy coding algorithms such as arithmetic coding~\cite{rissanen1979arithmetic, witten1987arithmetic} and transmitted as a sequence of bits. On the other side, the decoder recovers $\bm{\hat{y}}$ from the compressed codes, and subjects it to a parametric nonlinear synthesis transform $g_s(\bm{\hat{y}}; \bm{\phi_g})$ to build the reconstructed image $\bm{\hat{x}}$.
The nonlinear parametric transform $g_a(\bm{x}; \bm{\theta_g})$ and $g_s(\bm{x}; \bm{\phi_g})$ are implemented by convolutional neural networks and the parameters $\bm{\theta_g}$ and $\bm{\phi_g}$ are learned over a large amount of natural images.

The training goal is to minimize the expected length of the bitstream as well as the expected distortion of the reconstructed image with respect to the original image, giving rise to a rate-distortion optimization problem:
\begin{equation}
\begin{aligned}
    \centering
        & \text{minimize} \    R + \lambda \cdot D \\
        & R = \mathbb{E}_{\bm{x} \sim p_{x}}[-log_2 p_{\hat{y}}(Q(g_a(\bm{x})))] \\
        & D = \lambda \cdot \mathbb{E}_{\bm{x} \sim p_{x}}[d(\bm{x}, g_s(Q(g_a(\bm{x}))))]
    \label{eq:rd}
\end{aligned}
\end{equation}
where $\lambda$ is the Lagrange multiplier that determines the desired rate-distortion trade-off, $p_x$ is the (unknown) distribution of source images and $p_{\hat{y}}$ is a discrete entropy model.
$Q(\cdot)$ represents quantization, implemented as rounding to the nearest integer in most of the end-to-end compression papers. The first term (representing rate) corresponds to the cross entropy between the marginal distribution of the latent and the learned entropy model, which is minimized when the two distributions are identical. The second term (distortion) corresponds to the reconstructed error between the original image and the reconstructed image.


In most of the published end-to-end image compression CNNs, uniform scalar quantizer instead of vector quantizer (VQ) is adopted as the quantization function $Q(\cdot)$ in Eq.~\ref{eq:rd}, That is,
\begin{equation}
    \centering
    \hat{y}_i = Q(y_i) = \text{round}(y_i)
\end{equation}
where index $i$ runs over all elements of the vectors, including channels and spatial locations.
Note that both terms in Eq.~\ref{eq:rd} depend on the quantized values, and the derivatives of the quantization function are zero almost everywhere, causing gradient descent problem. To allow optimization via stochastic gradient descent, a relaxation technique is used to replace the quantizer with an additive i.i.d. uniform noise $\Delta y_i$, which has the same width as the quantization bins (one):
\begin{equation}
    \tilde{y}_i = y_i + \Delta{y}_i, \ \Delta{y}_i \in \mathcal{U}(-\frac{1}{2}, \frac{1}{2}).
\end{equation}

Very few articles~\cite{Agustsson2017_Soft,zhu2022unified} studied vector quantization in end-to-end image compression networks.
Agustsson~\textit{et~al.}~\cite{Agustsson2017_Soft} proposed a so-called soft-to-hard vector quantization module to bridge the nearest neighbor decision process of VQ and 
variational backpropagation that is necessary for the end-to-end CNN training. Their technique is a soft (continuous) relaxation of branching operation of VQ 
so that the VQ effects can be approximated in the end-to-end training.
Zhu~\textit{et~al.}~\cite{zhu2022unified} proposed a probabilistic vector quantization with cascaded estimation to estimate means and covariances.  However, the optimization goal of~\cite{zhu2022unified} only contains the distortion term, ignoring the vital design requirement of coding rates.



After the pioneering work of Ballé~\textit{et~al.}~\cite{Balle2017_End}, many other end-to-end compression methods have been proposed to further improve the R-D performances by introducing more complex nonlinear transforms and more efficient context entropy models.

Rippel~\textit{et~al.}~\cite{rippel2017,agustsson2019} proposed to learn the distribution of images using adversarial training to achieve better perceptual quality at extremely low bit rate.
Johnston~\textit{et~al.}~\cite{johnston2018} published a spatially adaptive bit allocation algorithm that efficiently uses a limited number of bits to encode visually complex image regions.
Li~\textit{et~al.}~\cite{li2018} developed a method to allocate the content-aware bit rate under the guidance of a content-weighted importance map.
Some papers focused on investigating the adaptive context model for entropy estimation to achieve a better trade-off between reconstruction errors and required bits (entropy), including~\cite{Mentzer2018_Cond,Balle2018_Vari,Minnen2018_Joint,Lee2019_Context},
among which the CNN methods of~\cite{Minnen2018_Joint,Lee2019_Context} are the first to outperform BPG in PSNR.
Choi~\textit{et~al.}~\cite{choi2019} published a novel variable-rate learned image compression framework with a conditional auto-encoder.
Cheng~\textit{et~al.}~\cite{Chen2020_Learned} proposed to use discretized Gaussian Mixture Likelihoods to parameterize the distributions of latent codes and achieved a more accurate and flexible entropy model.
Lin~\textit{et~al.}~\cite{lin2020spatial} proposed a novel spatial recurrent neural network for end-to-end image compression. The block based LSTM is utilized in spatial RNN to fully exploit spatial redundancy.
Minnen~\textit{et~al.}~\cite{minnen2020channel} introduced two enhancements,
channel-conditioning and latent residual prediction, that lead
to network architectures with better rate-distortion performance than existing context-adaptive models while minimizing serial processing. 
Zhang~\textit{et~al.} \cite{agdl} proposed a deep learning system for attention-guided
dual-layer image compression (AGDL) by introducing a novel idea of critical pixel set.

More recently, He~\textit{et~al.}~\cite{he2021checkerboard} proposed a parallelizable checkerboard context model to speed up the decoding process in the end-to-end image compression. Yang\textit{et~al.}~\cite{yang2021slimmable} proposed slimmable compressive autoencoders (SlimCAEs), where rate (R) and distortion (D) are jointly optimized for different capacities.
Gao~\textit{et~al.}~\cite{gao2021neural} proposed to conduct neural image compression via Attentional Multi-scale Back Projection and
Frequency Decomposition.
Kim~\textit{et~al.}~\cite{kim2022joint} proposed a novel entropy model called Information Transformer (Informer) that exploits both global and local information in a content-dependent manner using an attention mechanism.
He~\textit{et~al.}~\cite{he2022elic} proposed the uneven channel-conditional adaptive coding, motivated by the observation of energy compaction in learned image compression.
Cheng~\textit{et~al.}~\cite{cheng2022optimizing} proposed to optimize the image compression algorithm to be
noise-aware as joint denoising and compression to resolve the noisy image compression problem.
There are some other papers~\cite{shin2022expanded,pan2022content,li2022content} focused on the content adaptive approaches by updating the encoder-side component in the inference time to improve the coding performance.


\section{Design of LVQAC}
The main contribution of this work is to replace the uniform scalar quantization of feature values by the new LVQAC module outlined in the introduction, in the end-to-end design of compression CNNs.
The LVQAC module is composed of a lattice vector quantizer (LVQ) and a spatially adaptive companding (AC) mapping. LVQ, also known as poor man's VQ, is computationally almost as simple as scalar uniform quantization, but it enjoys higher coding efficiency thanks to more efficient spatial covering property of LVQ \cite{Conway}.


In addition, we couple LVQ with another compression technique of adaptive companding to improve the adaptability of LVQ to source statistics.

\subsection{Lattice Vector Quantization}
Let ${\{\bm{u}_1, ... , \bm{u}_n\}}$ be a set of linearly independent vectors in $\mathbb{R}^n$, which is the $n$-dimensional basis generating a lattice $\Lambda$.  $\Lambda$ consists of   
all points of the form
\begin{equation}
    \centering
    \bm{z} = \sum_{i=1}^{n} c_i \bm{u}_i,  \ c_i \in \mathbb{Z}.
\end{equation}
Let $\bm{U}=(\bm{u}_1, ... , \bm{u}_n)$ be the lattice generator matrix.
Any vector of the lattice can be represented as $\bm{z} = \bm{c} \bm{U}, \bm{c}=(c_1, ... , c_n)$, thus a lattice can be defined as:
\begin{equation}
    \centering
    \bm{z} \triangleq \{ \bm{z} \in \mathbb{R}^n: \bm{z} = \bm{c} \bm{U}, \bm{c} \in \mathbb{Z}^n \}.
\end{equation}
The Voronoi cell of a lattice point $\bm{z} \in \Lambda$ is defined as
\begin{equation}
    \centering
    V(\bm{z}) \triangleq \{ \bm{x} \in \mathbb{R}^n: ||\bm{x}-\bm{z}|| \leq ||\bm{x}-\hat{\bm{z}}||, \forall{\hat{\bm{z}} \in \Lambda} \}.
\end{equation}
The above structure is that
of lattice vector quantization~\cite{gray1998quantization,gersho2012vector}. The LVQ codewords are the centroids of the Voronoi polyhedrons which are of the same size and shape.

\begin{figure}[t]
    \centering
    \begin{subfigure}[b]{0.48\linewidth}
        \centering
        \includegraphics[width=\textwidth]{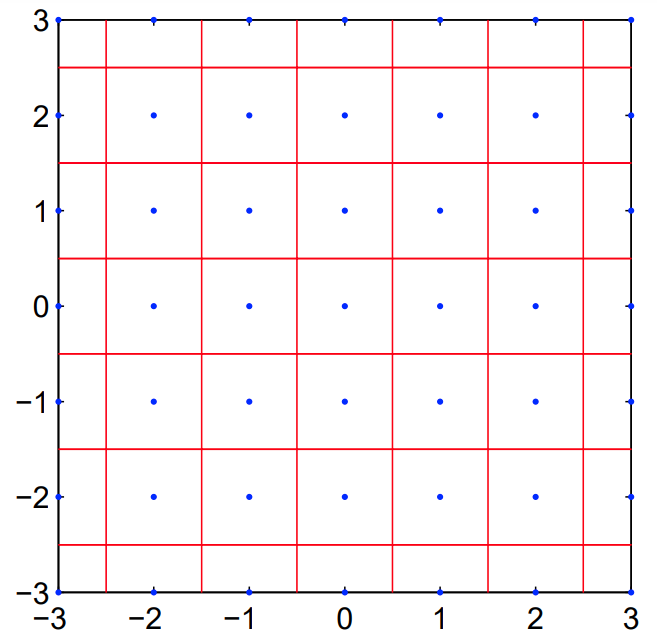}
        \caption{Square lattice $Z_2$.}
        \label{fig:square}
    \end{subfigure}
	\hfill
	\begin{subfigure}[b]{0.48\linewidth}
        \centering
        \includegraphics[width=\textwidth]{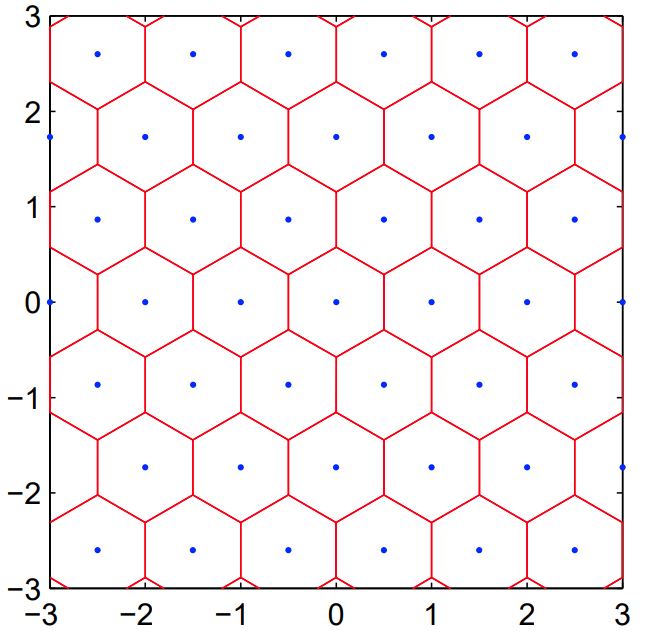}
        \caption{Hexagonal lattice $A_2$.}
        \label{fig:hexagonal}
    \end{subfigure}
    \caption{Two-dimensional square lattice $Z_2$ and hexagonal lattice $A_2$ and their Voronoi cells. }
    \label{fig:lattices_demo}
\end{figure}

Fig.~\ref{fig:lattices_demo} depicts the concept of LVQ.  It shows two examples: the two-dimensional square lattice (integer lattice) $Z_2$ and the hexagonal lattice $A_2$, together with their Voronoi cells.
The generator matrix for $Z_2$ lattice and $A_2$ lattice are as follows:
\begin{equation}
    \centering
    \bm{U}_{Z_2} =
    \begin{pmatrix}
        1 & 0 \\
        0 & 1
    \end{pmatrix},
    \qquad
    \bm{U}_{A_2} =
    \begin{pmatrix}
        1 & 0 \\
        1/2 & \sqrt{3}/2
    \end{pmatrix}.
\end{equation}
In fact, the square lattice $Z_2$ is equivalent to the simple uniform scalar quantizer. The hexagonal lattice $A_2$, however, offers a more efficient covering of 2D space than the square lattice $Z_2$.  Therefore, LVQ with the hexagonal lattice codebook offers higher coding efficiency than the square lattice codebook.
It is noteworthy that the hexagonal lattice is the optimal lattice for two-dimensional space~\cite{hexagonal} but it can not be directly applied to the situation of three or higher dimensional space.

\begin{figure*}[t]
    \centering
    \includegraphics[width=0.85\textwidth]{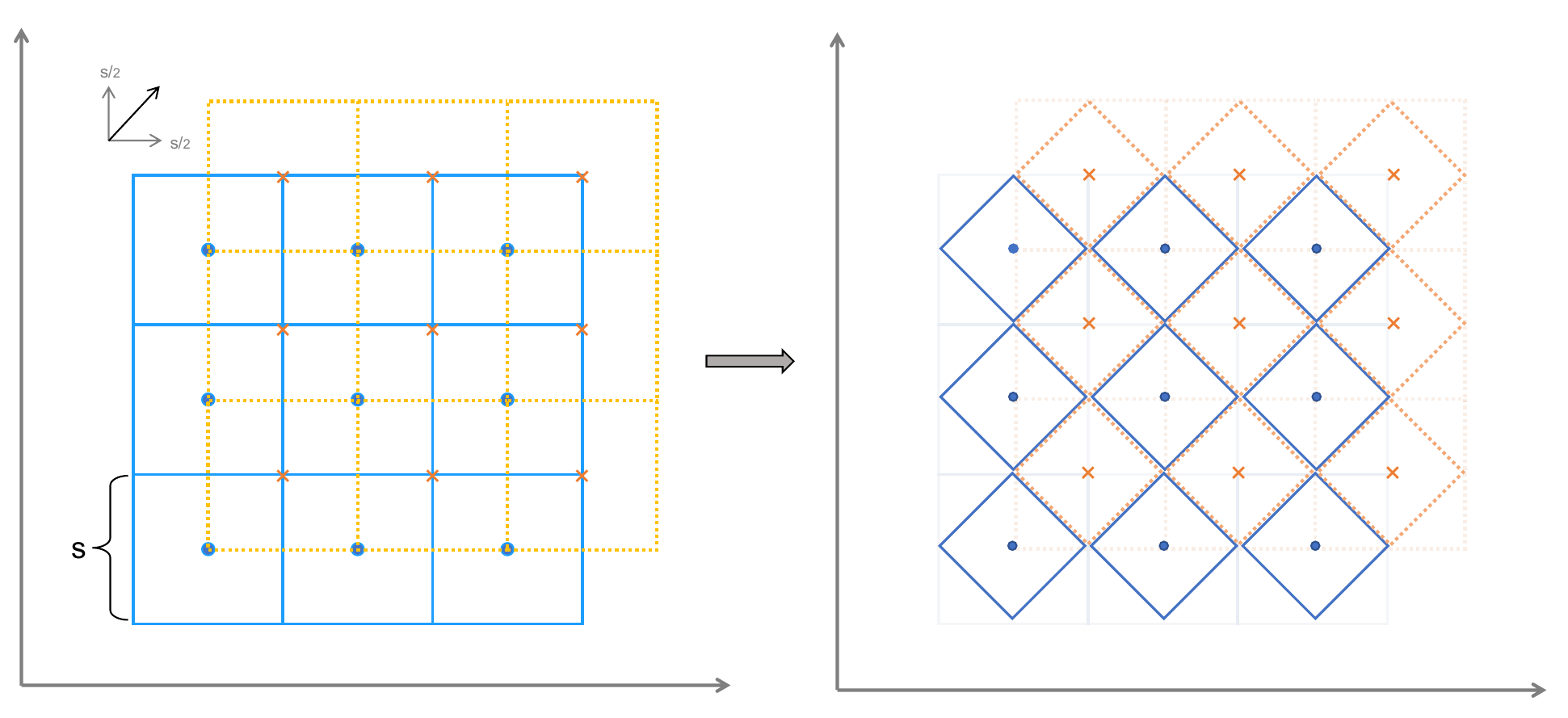}
    \caption{\textbf{Left}: Two codebooks used in diamond lattice quantization. The first codebook (blue one) is a square lattice and the second codebook (yellow one) is obtained by shifting the first codebook by $1/2$ quantization stride in each dimension.
    \textbf{Right}: Two-dimensional diamond lattice codebook. It can be obtained by shifting the square lattice like the left.}
    \label{fig:diamond}
\end{figure*}

For our task of quantizing CNN feature vectors, the dimensionality of the problem is much higher than two. In the interest of low complexity and dimension independence, we choose the diamond lattice (also known as body centred lattice) instead of more complex lattice structures.
The generator matrix of the $n$-dimensional diamond lattice 
is

\begin{equation}
    \centering
    \bm{U} =
    \begin{pmatrix}
        1   & 0      & 0    & 0 & \cdots & 0      & 0     \\
        1/2 & -1/2   & 0    & 0 & \cdots & 0      & 0     \\
        0   & 1/2    & -1/2 & 0 & \cdots & 0      & 0     \\
            & \vdots &      &   & \ddots & \vdots &       \\
        0   & 0      & 0    & 0 & \cdots & -1/2   & 0     \\
        0   & 0      & 0    & 0 & \cdots & 1/2    & -1/2  \\
    \end{pmatrix}.
\end{equation}
An example of two-dimensional diamond lattice is shown in Fig.~\ref{fig:diamond}.

The lattice vector quantization process can be decomposed into two scalar quantizations with respective codebooks plus one argmin operation.
The first codebook is a square lattice and the second codebook is obtained by shifting the first codebook by $1/2$ quantization stride in each dimension, as shown in Fig.~\ref{fig:diamond}.
Specifically, for $\bm{z}$ to be quantized, instead of doing the regular vector quantization (computing distances between $\bm{z}$ and all centroids and select the nearest one), we do twice scalar quantizations to each element of $\bm{z}$ based on two codebooks and get two quantized vectors $\bm{\hat{z}}_1$ and $\bm{\hat{z}}_2$. We choose the closer one as the lattice quantized result $\bm{\hat{z}}$, that is
\begin{equation}
    \centering
    \bm{\hat{z}}= \mathop{\arg\min}_{ \bm{q}=\{\bm{\hat{z}}_1, \bm{\hat{z}}_2 \} } \| \bm{z} - \bm{q} \|.
\end{equation}

As $\mathop{\arg\min}$ operation is not compatible with backpropagation, we instead use its soft relaxation which is proposed in \cite{Agustsson2017_Soft}.
The relaxation can be written as
\begin{equation}
    \centering
    \bm{\hat{z}} = \phi_a \bm{\hat{z}}_1 + \phi_b \bm{\hat{z}}_2.
\end{equation}
where the weights $(\phi_a, \phi_b)$ are obtained by applying softmax on the distances from $\bm{z}$ to $\bm{\hat{z}}_1$ and $\bm{\hat{z}}_2$,
that is:
\begin{equation}
    \centering
    (\phi_a, \phi_b) = \Psi(-\sigma[ \| \bm{z}-\bm{\hat{z}}_1 \|, \| \bm{z}-\bm{\hat{z}}_2 \| ]),
\end{equation}
where $\Psi$ is the softmax function, $\sigma$ controls the degree of relaxation (or called hardness of the soft assignment as in the original paper).

Specifically, given the latent representation $\bm{y} \in \mathbb{R}^{h \times w \times c}$ with $c$ channels and $h \times w$ size,
we apply the lattice quantization operation $\bm{Q_l}(\cdot)$
on the latent vector for each spatial location $\{ \bm{y}_j \in \mathbb{R}^c, 1 \leq j \leq hw \}$,
so the lattice quantized latent representation is:
\begin{equation}
    \centering
    \bm{\hat{y}} = \{ \bm{Q_l}(\bm{y}_j), \bm{y}_j \in \mathbb{R}^c, 1 \leq j \leq hw \}.
\end{equation}

It is noteworthy that in addition to transmitting the index of quantized representation,
we also need to transmit extra bits to identify the selected codebook besides the quantized vectors, but the bitrate of codebook selection is negligible.
As the lattice quantization is done to the latent vector for each spatial location, all elements in each spatial location share the same codebook. Thus the codebook only varies with spatial location.
For a given image of size ${h \times w \times 3}$,  the extracted latent representation will be of size
${\frac{h}{16} \times \frac{w}{16} \times c}$ and the codebook selection information will be of size ${\frac{h}{16} \times \frac{w}{16}}$ with 1-bit precision. The extra bitrate for codebook selection is less than 0.004 bpp even without binary compression.

\subsection{Spatially Adaptive Companding}
To further improve coding efficiency, we propose a spatially adaptive companding mapping to be coupled with the LVQ module to boost its adaptation to source statistics. 
Instead of adopting a universal companding to all elements over channels and spatial locations, we 
choose to learn different companding functions for different spatial locations.
To reduce the difficulty of learning and also guarantee the inference speed, we restrict the companding to the form of A-law function.
For a given input $x$, the equation for A-law encoding is as follows:

\begin{equation}
    \centering
    F(x) = \text{sgn}(x)
    \left\{
        \begin{array}{ll}
        \dfrac{A|x|}{1+\text{ln}(A)}, & |x| < \dfrac{1}{A} \\ 
        \\
        \dfrac{1+ln(A|x|)}{1+\text{ln}(A)}, & \dfrac{1}{A} \leq |x| \leq 1 
        \end{array}
    \right .
\end{equation}

where $A$ is the scale parameter. 
A-law decoding is given by the inverse function:

\begin{equation}
    \centering
    F^{-1}(y) = \text{sgn}(y)
    \left\{
        \begin{array}{ll}
        \dfrac{|y|(1+\text{ln}(A))}{A}, & |y| < \dfrac{1}{1+\text{ln}(A)} \\ 
        \\
        \dfrac{e^{-1+|y|(1+\text{ln}(A))}}{A}, & \dfrac{1}{1+\text{ln}(A)} \leq |x| \leq 1 
        \end{array}
    \right .
\end{equation}

The A-law function and its inverse are both continuous and derivative, which implies that they can be embedded into the end-to-end optimization system.

Specifically, we learn different scale parameters $A$ for different spatial locations and apply the unique learned A-law companding to all elements at the same location. Learning of scale parameters $A$ is implemented by a shallow convolution network $h$, which consits of several convolution layers. It takes the latent representation $y$ as input; and predicts the optimal scale parameters of A-law functions for different spatial locations, $\bm{a} = h(\bm{y})$. Then we apply these learned A-law companding functions $f(F; \bm{a})$ on the latent representations to scale it to the ideal dynamic range for better quantization performance. In the other side, after decoding $\hat{\bm{y}}$ from bitstream, the inverse A-law companding $f^{-1}(F^{-1}; \bm{a})$ is applied on $\hat{\bm{y}}$ to scale it to the original dynamic range.

Given the proposed lattice vector quantization module and spatially adaptive companding mapping, the optimization goal should be rewritten to:  
\begin{equation}
\begin{aligned}
    & \text{minimize} \    R + \lambda \cdot D \\
    & R = \mathbb{E}_{\bm{x} \sim p_{x}}[-log_2 p_{\hat{y}}(\bm{Q_l}(\bm{f}(g_a(\bm{x}))))] \\ 
    & D = \lambda \cdot \mathbb{E}_{\bm{x} \sim p_{x}}
        [d(\bm{x}, g_s(\bm{f}^{-1}(\bm{Q_l}(\bm{f}(g_a(\bm{x})))))]
    \label{eq:rd2}
\end{aligned}
\end{equation}
where $\bm{Q_l}$ is the proposed LVQ module, $\bm{f}$ and $\bm{f}^{-1}$ are the learned spatially adaptive A-law function and its inverse.

\section{Experiments}
We conduct extensive experiments to systematically evaluate and analyze the effectiveness of the proposed LVQAC scheme.
Specifically, we select two representative end-to-end CNN compression architectures Cheng2020~\cite{Chen2020_Learned} and Minnen2018~\cite{Minnen2018_Joint} and replace their uniform scalar quantizers by the proposed LVQAC module to demonstrate the superiority of LVQAC.
We show the R-D performance comparisons between the models with uniform scalar quantization and LVQAC for each architecture, and also provide visual comparisons.
Besides, the encoding and decoding latencies are tabulated to validate the computational efficiency of LVQAC.
Other analyses such as an ablation study on the effects of each component of LVQAC are further given.

\subsection{Dataset and Training Details}
The training dataset consists of the largest 8000 images picked from ImageNet~\cite{imagenet}, following the previous work~\cite{he2022elic}. 
Training images are then random-cropped to $512 \times 512$ and batched into $16$.
We train each model using the Adam optimizer with $\beta-1=0.9, \beta_2=0.999$. 
Each model is trained with an initial learning rate $10^{-4}$ for 2000 epochs (1M iterations), and then we decay the learning rate to $10^{-5}$ for another 100-epoch training. 
Tests involve two image sets: Kodak (24 images) and CLIC Professional valid set (41 images).
The network and the proposed LVQAC module are both implemented in PyTorch~\cite{pytorch} and CompressAI~\cite{begaint2020compressai}.
All experiments are conducted with four RTX 2080ti GPUs.

\begin{figure*}[t]
    \centering
    \begin{subfigure}[b]{0.49\textwidth}
        \centering
        \includegraphics[width=\textwidth]{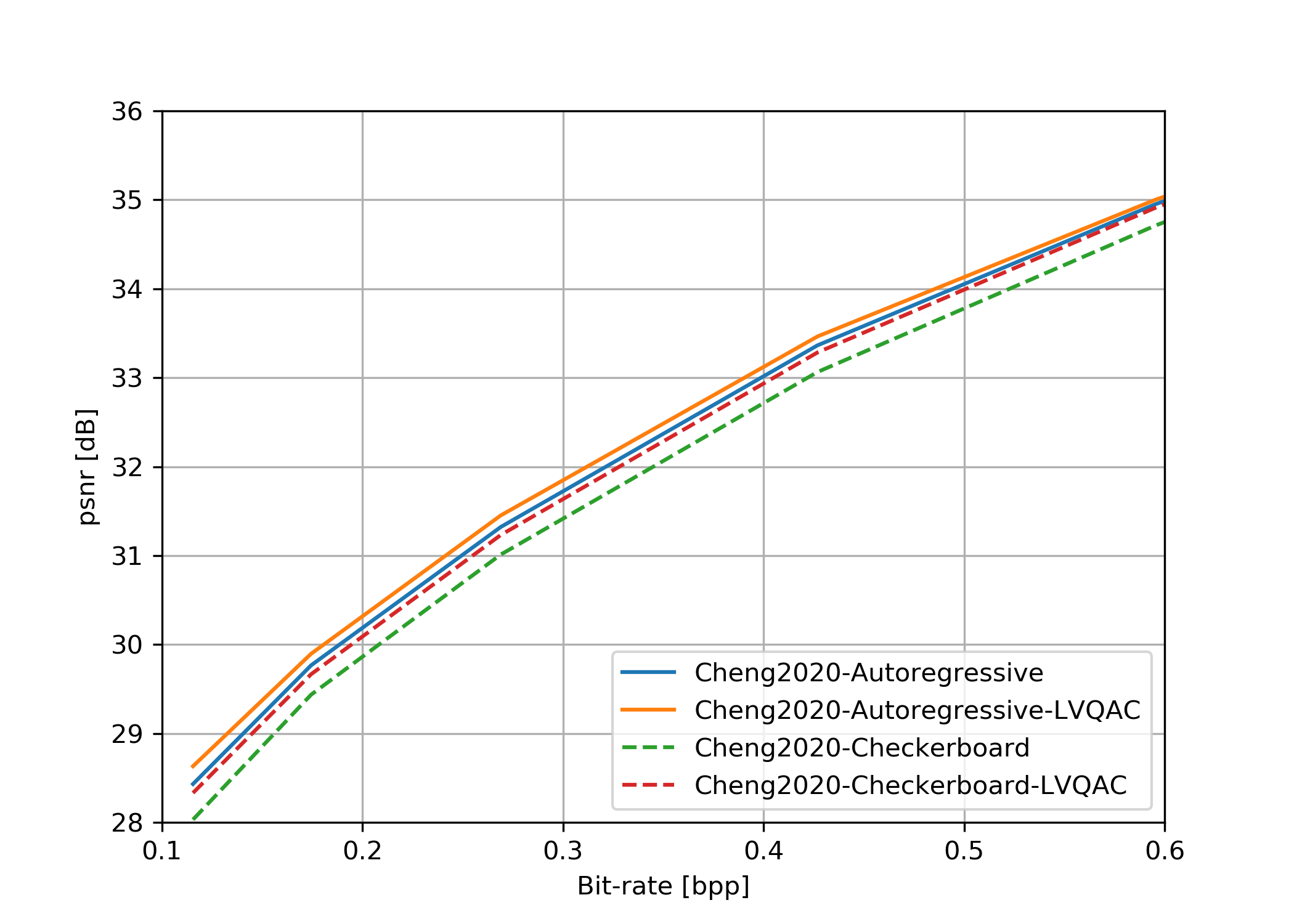}
        \caption{Cheng2020~\cite{Chen2020_Learned}, Kodak.}
        \label{fig:cheng2020_kodak_psnr}
    \end{subfigure}
	\hfill
	\begin{subfigure}[b]{0.49\textwidth}
        \centering
        \includegraphics[width=\textwidth]{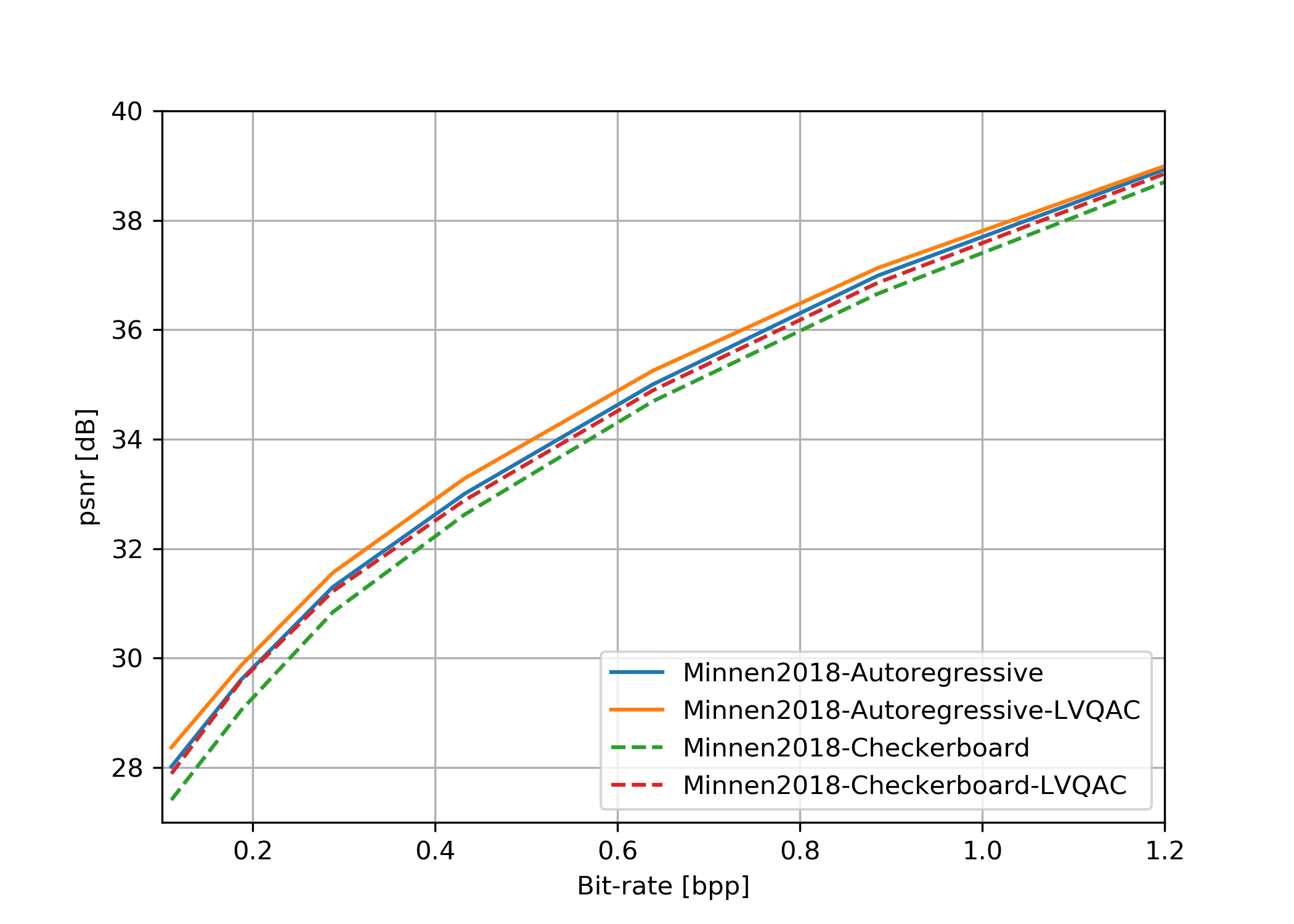}
        \caption{Minnen2018~\cite{Minnen2018_Joint}, Kodak.}
        \label{fig:minnnen2018_kodak_psnr}
    \end{subfigure}
    \caption{Rate-distortion curves on Kodak dataset. All models are optimized for MSE.}
    \label{fig:kodak_rd}
    \vspace{-0.5cm}
\end{figure*}
\begin{figure*}[t]
    \begin{subfigure}[b]{0.49\textwidth}
        \centering
        \includegraphics[width=\textwidth]{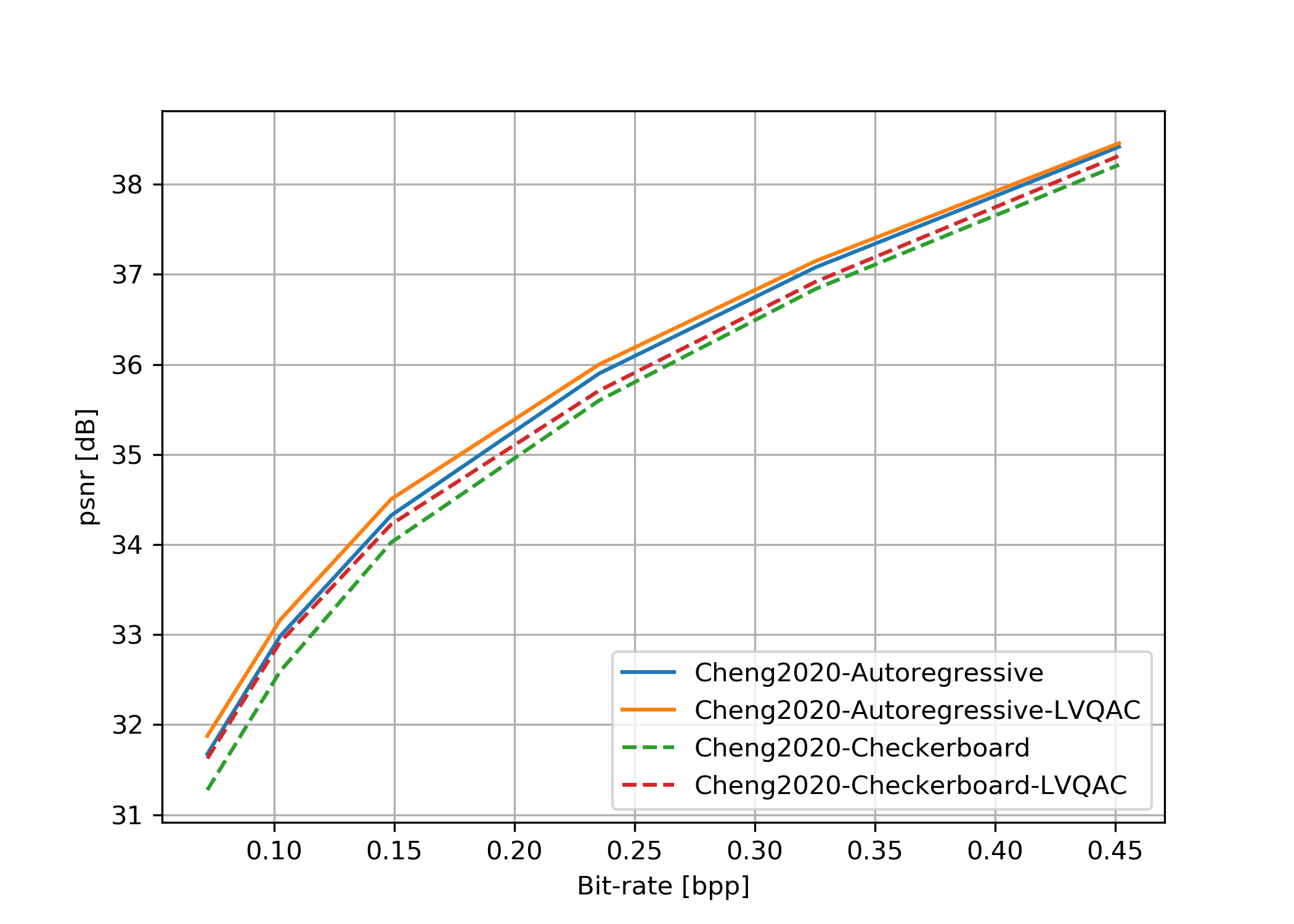}
        \caption{Cheng2020~\cite{Chen2020_Learned}, CLIC Pro.}
        \label{fig:cheng2020_clicpro_psnr}
    \end{subfigure}
	\hfill
	\begin{subfigure}[b]{0.49\textwidth}
        \centering
        \includegraphics[width=\textwidth]{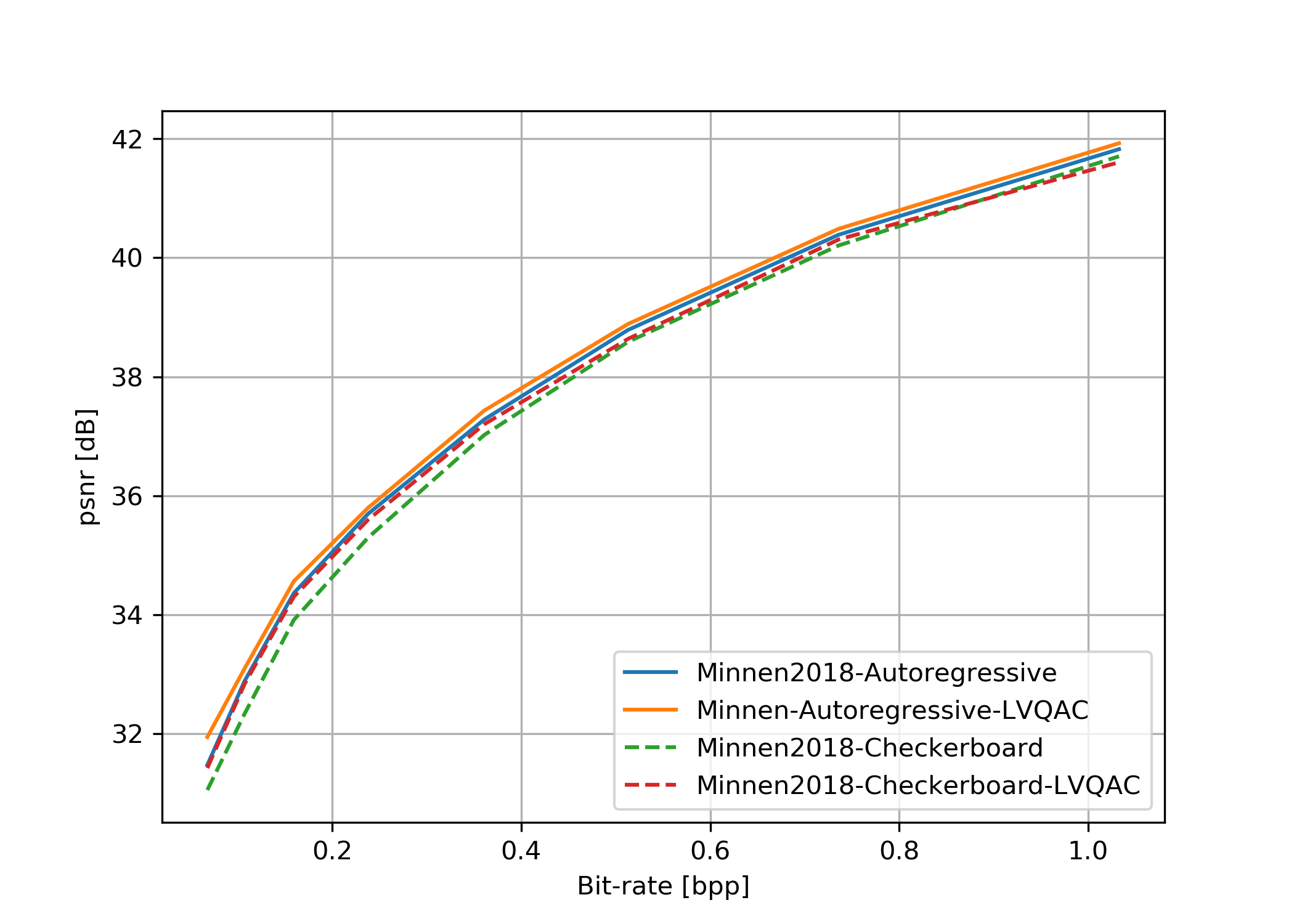}
        \caption{Minnen2018~\cite{Minnen2018_Joint}, CLIC Pro. }
        \label{fig:minnnen2018_clicpro_psnr}
    \end{subfigure}
    \caption{Rate-distortion curves on CLIC Pro dataset. All models are optimized for MSE.}
    \label{fig:clicpro_rd}
    \vspace{-0.2cm}
\end{figure*}

\subsection{Rate-Distortion Performance}
We evaluate our LVQAC scheme based on two representative published architectures
Cheng2020~\cite{Chen2020_Learned} and Minnen2018~\cite{Minnen2018_Joint}.
The reason for choosing these two architectures is that they are the milestones in the end-to-end optimized image compression area, as Minnen2018~\cite{Minnen2018_Joint} is the first learning-based method to outperform BPG in PSNR metric and Cheng2020~\cite{Chen2020_Learned}  is the first work to achieve comparable performance with latest compression standard Versatile Video Coding (VVC) regarding PSNR.  
These two architectures were originally proposed with an autoregressive (serial) context model. 
To fully explore the potentiality of LVQAC, we consider the two architectures with two different context models, namely the autoregressive context model and checkerboard context model~\cite{he2021checkerboard}.
For each architecture, given two different context models, we do not change any other architectures except by replacing the uniform scalar quantizer with the proposed LVQAC scheme.
We implement the two representative models coupled with uniform scalar quantizer and LVQAC, respectively. 
The Lagrange multipliers are set to $\lambda = \{ 0.0016, 0.0032,0.0075, 0.015, 0.03, 0.045 \}$.

Fig.~\ref{fig:kodak_rd} and Fig.~\ref{fig:clicpro_rd} graphically
presents the rate-distortion behaviors of different methods on Kodak and CLIC Pro datasets 
in PSNR metric.
We can see that for any architecture, no matter with autoregressive or checkerboard context model, replacing uniform quantizer by LVQAC achieves better rate-distortion performance. 
Besides, LVQAC brings greater performance gain when coupled with the checkerboard context model than being coupled with the autoregressive context model. 
The CNN model adopted LVQAC + checkerboard context model can achieve a similar performance as the counterpart CNN using a far more expensive autoregressive context model + uniform scalar quantizer, yet the former is much lighter and faster than the latter. We show the detailed computational complexity of LVQAC in subsection~\ref{subsec:latency}.
We can also find that LVQAC can bring more performance gain at the low bit rate than at the high bit rate. 
This is expected as uniform quantization can approach the rate-distortion optimality 
at very high bit rate~\cite{gersho2012vector}.

\begin{figure*}[t]
    \centering
    \begin{subfigure}[b]{0.24\linewidth}
        \centering
        \includegraphics[width=\textwidth, height=0.7\textwidth]{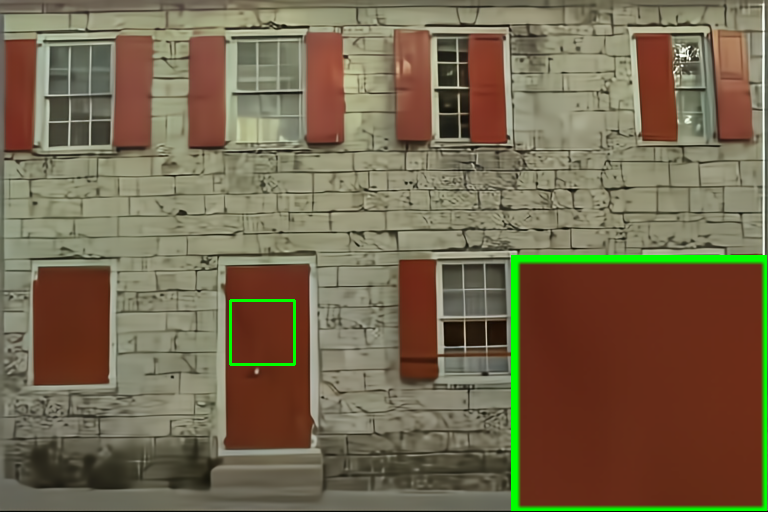}
        \caption{BPG}
    \end{subfigure}
	\hfill
	\begin{subfigure}[b]{0.24\linewidth}
        \centering
        \includegraphics[width=\textwidth, height=0.7\textwidth]{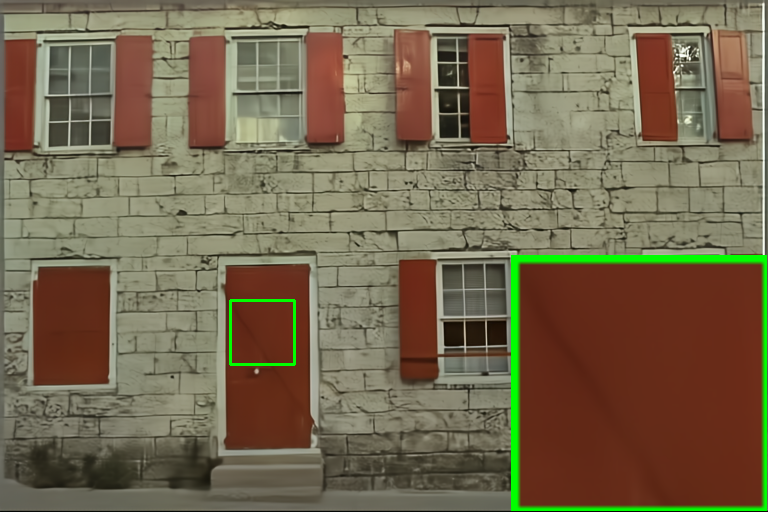}
        \caption{Cheng2020, Scalar}
    \end{subfigure}
	\hfill
    \begin{subfigure}[b]{0.24\linewidth}
        \centering
        \includegraphics[width=\textwidth, height=0.7\textwidth]{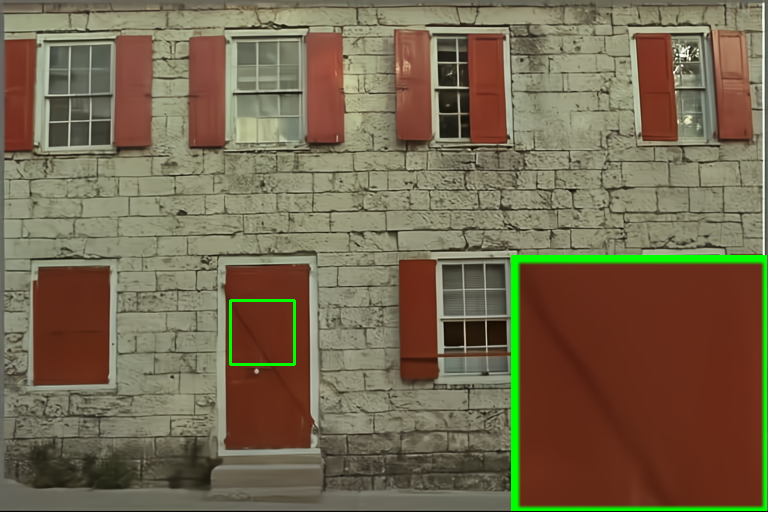}
        \caption{Cheng2020, LVQAC}
    \end{subfigure}
	\hfill
    \begin{subfigure}[b]{0.24\linewidth}
        \centering
        \includegraphics[width=\textwidth, height=0.7\textwidth]{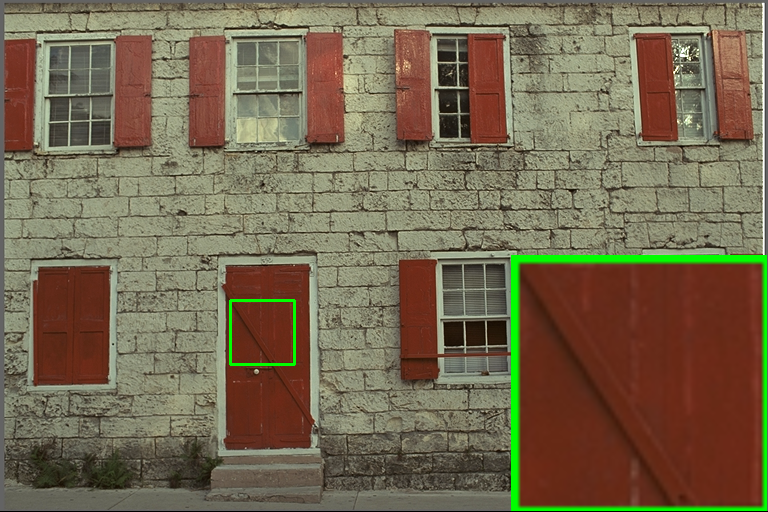}
        \caption{Ground Truth}
    \end{subfigure}
    \caption{Visual comparisons of different methods on 'kodim01.png' from Kodak dataset. }
    \label{fig:kodim01}
\end{figure*}
\begin{figure*}[!h]
    \centering
    \begin{subfigure}[b]{0.24\linewidth}
        \centering
        \includegraphics[width=\textwidth, height=0.7\textwidth]{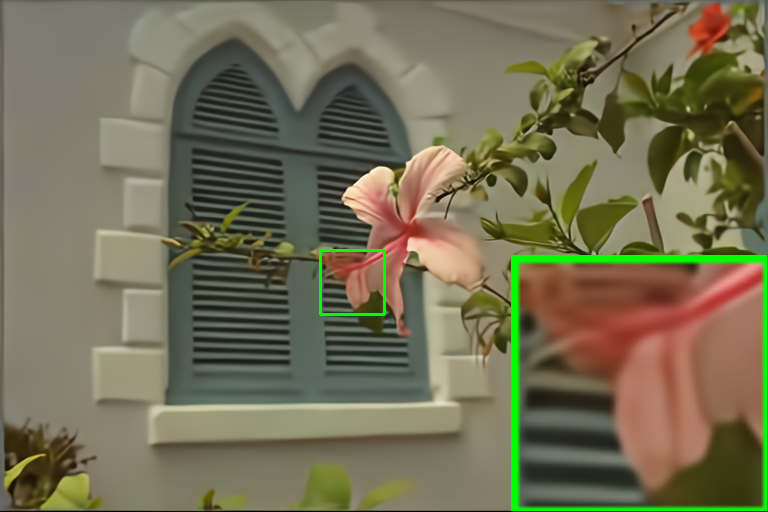}
        \caption{BPG}
    \end{subfigure}
	\hfill
	\begin{subfigure}[b]{0.24\linewidth}
        \centering
        \includegraphics[width=\textwidth, height=0.7\textwidth]{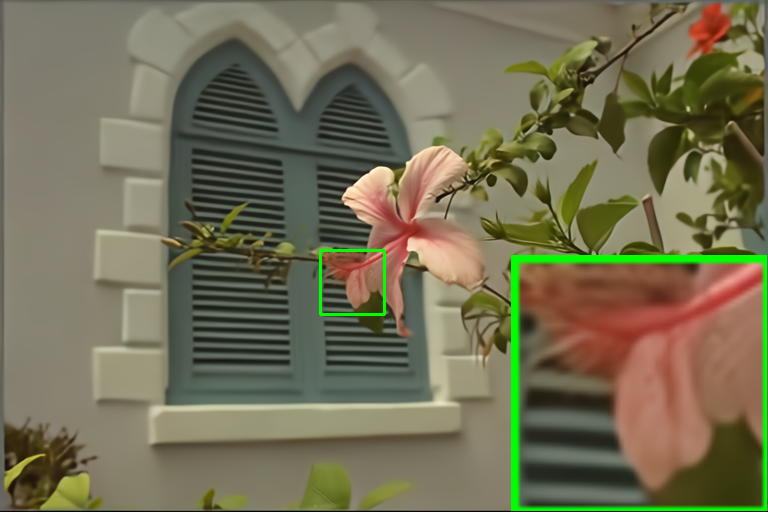}
        \caption{Cheng2020, Scalar}
    \end{subfigure}
	\hfill
    \begin{subfigure}[b]{0.24\linewidth}
        \centering
        \includegraphics[width=\textwidth, height=0.7\textwidth]{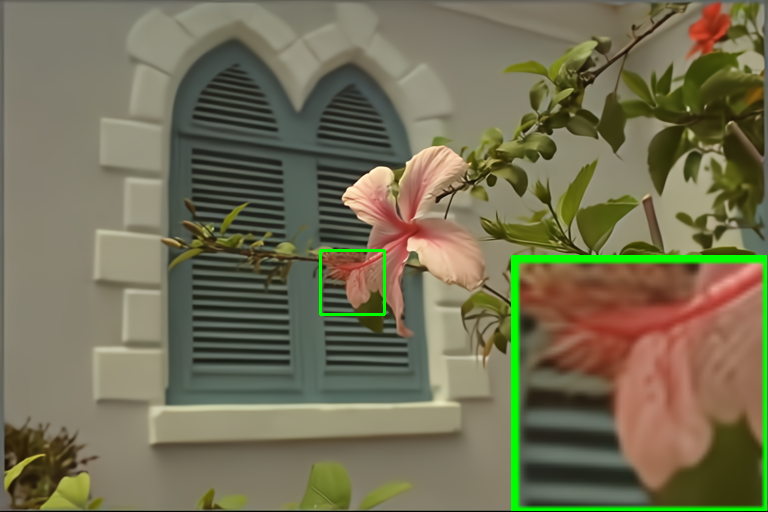}
        \caption{Cheng2020, LVQAC}
    \end{subfigure}
	\hfill
    \begin{subfigure}[b]{0.24\linewidth}
        \centering
        \includegraphics[width=\textwidth, height=0.7\textwidth]{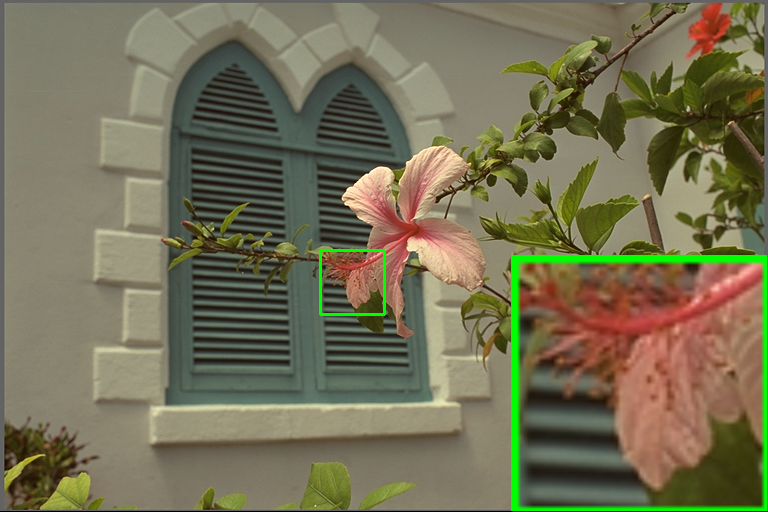}
        \caption{Ground Truth}
    \end{subfigure}
    \caption{Visual comparisons of different methods on 'kodim07.png' from Kodak dataset.  }
    \label{fig:kodim07}
\end{figure*}
\begin{figure*}[!h]
    \centering
    \begin{subfigure}[b]{0.24\linewidth}
        \centering
        \includegraphics[width=\textwidth]{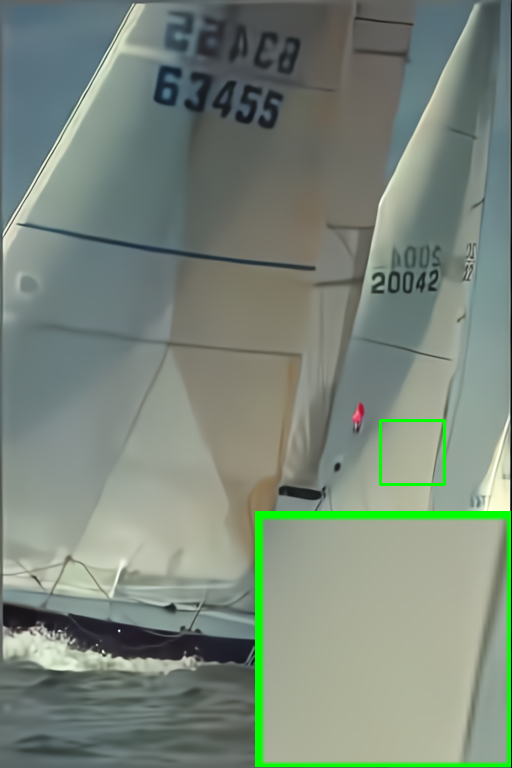}
        \caption{BPG}
    \end{subfigure}
	\hfill
	\begin{subfigure}[b]{0.24\linewidth}
        \centering
        \includegraphics[width=\textwidth]{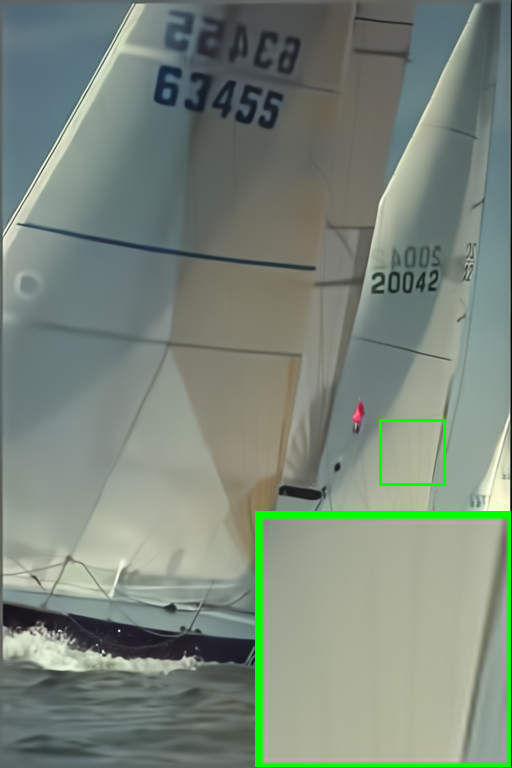}
        \caption{Cheng2020, Scalar}
    \end{subfigure}
	\hfill
    \begin{subfigure}[b]{0.24\linewidth}
        \centering
        \includegraphics[width=\textwidth]{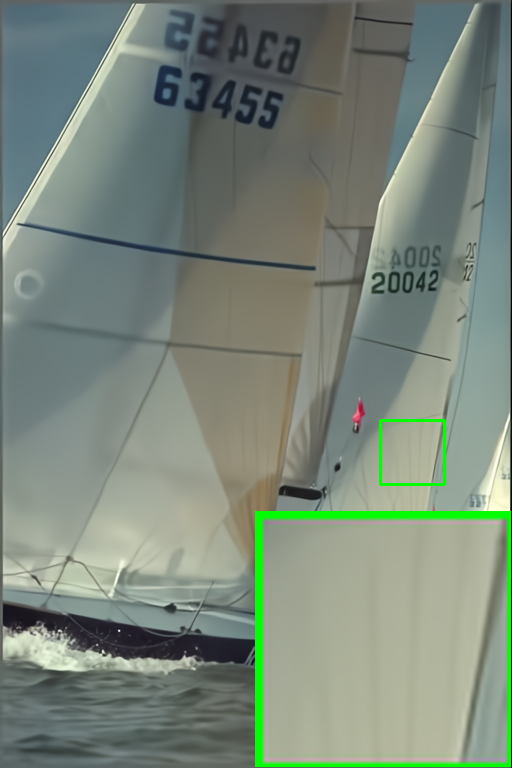}
        \caption{Cheng2020, LVQAC}
    \end{subfigure}
	\hfill
    \begin{subfigure}[b]{0.24\linewidth}
        \centering
        \includegraphics[width=\textwidth]{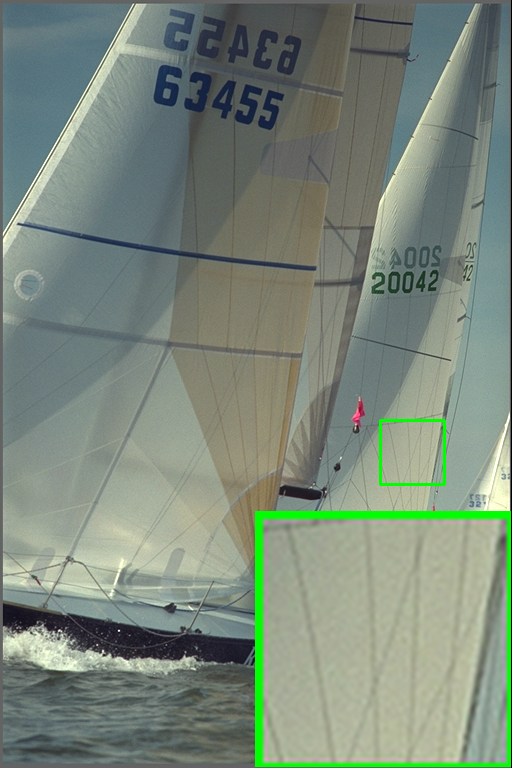}
        \caption{Ground Truth}
    \end{subfigure}
    \caption{Visual comparisons of different methods on 'kodim10.png' from Kodak dataset.  }
    \label{fig:kodim10}
\end{figure*}

\subsection{Visual Comparison}
We also provide visual comparisons to demonstrate the effectiveness of the proposed LVQAC scheme.
Fig.~\ref{fig:kodim01}, \ref{fig:kodim07} and \ref{fig:kodim10} show the visual results of different methods, 
including BPG, Cheng2020 with scalar quantizer, Cheng2002 with LVQAC and the Ground Truth. 
We pick three images 'kodim01.png', 'kodim07.png' and 'kodim10.png' from Kodak dataset to show the comparisons.
The bit rates of all methods are adjusted to the same level for the fairness.
In Fig.~\ref{fig:kodim01}, LVQAC can preserve the sharper edges on doors than uniform scalar quantizer.
In Fig.~\ref{fig:kodim07}, LVQAC also achieves richer and clearer textures on flowers over uniform scalar quantizer.
Another interesting observation is that the faint lines in Fig.~\ref{fig:kodim10} are both removed by BPG and scalar quantizer, but LVQAC can preserve such features.
These observations reveal that the end-to-end CNN compression model coupled with LVQAC can produce much sharper edges and clearer textures than the counterpart CNN coupled with scalar quantizer.
We present more high-resolution visual examples in the supplementary material.

\begin{table*}[!ht]
    \centering
    \caption{BD-Rates and inference times of different methods. The BD-Rate data is calculated relative to VVC (YUV 444) from PSNR-BPP curve on Kodak. [A] represnts the autoregressive context model, [C] represents the checkerboard context model.}
    \label{tab:latency}
    \renewcommand\arraystretch{1.2}
    \begin{tabular}{clccc}
        \hline
        \textbf{Architecture} & \textbf{Context + Quantizer} & \textbf{BD-Rate (\%)} & 
        \textbf{Enc. Time (ms)} & \textbf{Dec. Time (ms)} \\
        \hline
        \multirow{4}{*}{Cheng2020\cite{Chen2020_Learned}} 
            & \quad [A] + LVQAC  & 3.12 & 48.65 & $>10^3$ \\
            & \quad [A] + Scalar & 3.35 & 41.42 & $>10^3$ \\
            \cline{2-5}
            & \quad [C] + LVQAC    & 3.39 & 52.66 & 55.84   \\
            & \quad [C] + Scalar   & 3.89 & 45.43 & 53.01   \\
        \hline
        \multirow{4}{*}{Minnen2018\cite{Minnen2018_Joint}} 
            & \quad [A] + LVQAC  & 14.75 & 25.87 & $>10^3$ \\
            & \quad [A] + Scalar & 14.92 & 18.14 & $>10^3$ \\
            \cline{2-5}
            & \quad [C] + LVQAC    & 16.17 & 26.23 & 27.10   \\
            & \quad [C] + Scalar   & 20.00 & 18.84 & 24.52   \\
        \hline
    \end{tabular}
    \vspace{-0.1cm}
\end{table*}

\subsection{Encoding and Decoding Latency}
\label{subsec:latency}
We provide a detailed analysis of encoding and decoding latency in this subsection.
Specifically, we evaluate the inference latency of end-to-end CNN compression models with different quantizers: uniform scalar quantizer and the proposed LVQAC.
Table~\ref{tab:latency} tabulate the encoding time and decoding time of different architectures coupled with scalar quantizer and LVQAC. 
We can see that, for the architecture Cheng2020 with checkerboard context, LVQAC only increases the encoding time by about 15\% and the decoding time by about 5\%, yet improves the performance to the same level as that using autoregressive context. It is noteworthy that the decoding time of the model using autoregressive context is two orders of magnitude higher than that of the checkerboard context.
For the architecture Cheng2020 with autoregressive context model, the increase of encoding time is about 17\%, and the increase of decoding time is negligible due to the huge base ($>10^3$).
Similar observations can be obtained on the Minnen2018 architecture.
These observations reveal that 
replacing uniform quantizer by LVQAC can achieve better rate-distortion performance without significantly increasing the model complexity; the simpler the context-sensitive entropy model, the greater the performance gain.

\subsection{Ablation Study}
To investigate the impacts of the proposed LVQAC scheme, we conduct an ablation study to investigate the importance of 
each component in the proposed LVQAC scheme. We experiment progressively to explore the respective contribution of the lattice vector quantization module and spatially adaptive companding module.  
Specifically, we progressively add LVQ and AC modules into the network and test the coding performance.
Table~\ref{tab:ablation} shows the results of the ablation study. It can be seen that lattice vector quantization and spatially adaptive companding contribute almost equally to the performance gain.

\begin{table}[t]
    \centering
    \caption{Results of the progressive ablation study.
    The BD-Rate data is calculated relative to VVC (YUV 444) from PSNR-BPP curve on Kodak. 
    [A] represnts the autoregressive context model, [C] represents the checkerboard context model.}
    \label{tab:ablation}
    \renewcommand\arraystretch{1.2}
    \begin{tabular}{clc}
        \hline
        \textbf{Architecture} & \textbf{Context + Quantizer} & \textbf{BD-Rate} \\
        \hline
        \multirow{6}{*}{Cheng2020\cite{Chen2020_Learned}}
            & \quad [A] + LVQ + AC & 3.12  \\
            & \quad [A] + LVQ      & 3.24  \\
            & \quad [A] + Scalar + AC  & 3.30 \\
            & \quad [A] + Scalar   & 3.35  \\ 
            \cline{2-3}
            & \quad [C] + LVQ + AC & 3.39  \\
            & \quad [C] + LVQ      & 3.68  \\
            & \quad [C] + Scalar + AC  & 3.82  \\
            & \quad [C] + Scalar   & 3.89  \\
        \hline
        \multirow{6}{*}{Minnen2018\cite{Minnen2018_Joint}} 
            & \quad [A] + LVQ + AC & 14.75  \\
            & \quad [A] + LVQ      & 14.86  \\
            & \quad [A] + Scalar + AC   & 14.88 \\
            & \quad [A] + Scalar   & 14.92  \\
            \cline{2-3}
            & \quad [C] + LVQ + AC & 16.17  \\
            & \quad [C] + LVQ      & 18.43  \\
            & \quad [C] + Scalar + AC   & 19.29  \\
            & \quad [C] + Scalar   & 20.00  \\
        \hline
    \end{tabular}
    \vspace{-0.2cm}
\end{table}

\subsection{Limitation}
The main limitation of this work relies on that it is difficult to know the optimal lattice structure for any given dimension. In this paper, we choose the diamond lattice structure due to its simplicity and easy extension to any dimension. However, even though diamond lattice is not the optimal lattice structure in the vast majority of cases, 
it is still much more efficient than uniform scalar quantizer without significantly increasing the model complexity.
Another limitation of LVQAC is that if the context model is complex and powerful enough, such as an autoregressive context, then the gain brought by LVQAC will be very limited. Thus the application of LVQAC may be restricted to those lightweight end-to-end CNN compression models.

\section{Conclusion and Future Work}
In this paper, we present a novel Lattice Vector Quantization scheme coupled with a spatially Adaptive Companding (LVQAC) mapping.
LVQ can better exploit the inter-feature dependencies than scalar uniform quantization while being computationally almost as simple as the latter.
Moreover, to improve the adaptability of LVQ to source statistics, we couple a spatially adaptive companding (AC) mapping with LVQ.
The resulting LVQAC design can be easily embedded into any end-to-end optimized image compression system.
Extensive experiments demonstrate that for any end-to-end CNN image compression models, replacing uniform quantizer by LVQAC achieves better rate-distortion performance without significantly increasing the model complexity.
Future work will focus on exploring the optimal lattice vector quantizer for a given dimension and how to embed it into the end-to-end image compression system.

{
\small
\bibliographystyle{ieee_fullname}
\bibliography{lvqac}
}

\end{document}